\documentclass[a4paper,aps,superscriptaddress,floatfix,nofootinbib,twocolumn]{revtex4-1}

\usepackage{mathbbol}
\usepackage[pdftex]{graphicx}
\usepackage{latexsym,amsmath,verbatim,amssymb,txfonts}
\usepackage{color}
\usepackage{rotating}
\usepackage{verbatim}
\usepackage{multirow}
\usepackage[english]{babel}
\usepackage{comment}
\usepackage{hyperref}
\usepackage{dsfont}

\newcommand{\SM}{~\cite{SMat}}

\begin{document}

\title{Influence maximization on temporal networks}

\author{\c{S}irag Erkol}
\affiliation{Center for Complex Networks and Systems Research, Luddy School
  of Informatics, Computing, and Engineering, Indiana University, Bloomington,
  Indiana 47408, USA}

\author{Dario Mazzilli}
\affiliation{Center for Complex Networks and Systems Research, Luddy School
  of Informatics, Computing, and Engineering, Indiana University, Bloomington,
  Indiana 47408, USA}
  
\author{Filippo Radicchi}
\affiliation{Center for Complex Networks and Systems Research, Luddy School
  of Informatics, Computing, and Engineering, Indiana University, Bloomington,
  Indiana 47408, USA}
\email{filiradi@indiana.edu}

\begin{abstract}
We consider the optimization problem of seeding a spreading process on a temporal network so that the expected size of the resulting outbreak is maximized. We frame the problem for a spreading process following the rules of the susceptible-infected-recovered model with temporal scale equal to the one characterizing the evolution of the network topology.
We perform a systematic analysis based on a corpus of 12 real-world temporal networks and quantify the performance of solutions to the influence maximization problem obtained using different level of information about network topology and dynamics. We find that having perfect knowledge of the network topology but in a static and/or aggregated form is not helpful in solving the influence maximization problem effectively. Knowledge, even if partial, of the early stages of the network dynamics appears instead essential for the identification of quasioptimal sets of influential spreaders.
\end{abstract}

\maketitle

%%%%%%%%%%%%%%%%%%%%%%%%%%%%%%%%%%%%%%%%%%%%%%%%%%%%%%%%%%%%%%%%%%%%%%%%%%%%%%%%%%%%%%%%%%%%%%%%%%%%%%%%%%%%%%%%%%%%%%%%%%%%%%%%%%%%%%%%%%%%%%%%%%%%%%%%%%%%%%%%%%%%%%%%

\section{Introduction}

Influence maximization is a classical optimization problem in network science~\cite{domingos2001mining, kempe2003maximizing}. The problem consists in finding the set of initial spreaders that maximizes the expected outbreak size of a spreading process occurring on a network. The problem is generally solved for a fixed size of the set of initial spreaders. The size of the set is small if compared to the one of the network but large enough to forbid the use of brute-force algorithms for finding exact solutions to the problem. Most of the research on the topic is devoted to the development of approximate algorithms aiming at finding good solutions in a computationally feasible manner. Examples include greedy optimization techniques with relatively high computational complexity but guaranteed performance that can be applied to medium-size networks~\cite{kempe2003maximizing, leskovec2007cost, chen2009efficient, goyal2011celf++, nguyen2016stop, hu2018local},
and a multitude of approximate algorithms based on network centrality metrics applicable to large-scale graphs, e.g., Refs.~\cite{estrada2005subgraph,kitsak2010identification, de2014role, morone2015influence, lu2016vital, clusella2016immunization, zdeborova2016fast}. For a systematic analysis of several methods for the identification of influential spreaders in networks see Ref.~\cite{erkol2019systematic}.

Influence maximization has been traditionally studied on static graphs. However, several real-world networks display nontrivial edge temporal variability~\cite{holme2012temporal}. If structural variations happen on a timescale comparable with the one of the spreading dynamics, then the two processes interact in a highly nontrivial manner~\cite{prakash2010virus, karsai2011small, perra2012activity, valdano2015analytical}. Most of the work in the area of spreading processes on time-varying networks has been focusing on the characterization of their critical properties. Some attention has been devoted to the problem of influence maximization. Specifically, Habiba and Berger-Wolf~\cite{habiba2007maximizing} extend the work by Kempe \emph{et al.}~\cite{kempe2003maximizing} to temporal networks for the susceptible-infected (SI) and linear threshold (LT) models. In their modeling framework, the optimization problem consists in maximizing of a submodular function, thus appearing similar to the one valid for static graphs. However, they show also that ignoring the time order of the interactions generate results that do not relate to the ground-truth spreading process taking place on the temporal network. Similar conclusions are reached by Osawa \emph{et al.}, who study the problem of influence maximization on temporal networks for the SI model~\cite{osawa2015selecting}. Michalski \emph{et al.} model the temporal network as a multilayer network and the spreading dynamics using the LT model~\cite{michalski2014seed}. They analyze solutions to the influence maximization problem under different granularities for the temporal network, including time-aggregated versions of the network. Murata \emph{et al.} propose heuristic methods to solve the influence maximization problem on temporal networks~\cite{murata2018extended}. Han \emph{et al.}~\cite{han2017influence} and Zhuang \emph{et al.}~\cite{zhuang2013influence} propose a method where it is assumed that only the topology of the first layer of a temporal network is known. The topology of the succeeding temporal layers can only be discovered based on partial probing of nodes in the network, and such partial topological information is used to select influential spreaders. Gayraud \emph{et al.} focus on the independent cascade (IC) and LT models in a theoretical study about the properties of the influence maximization problem~\cite{gayraud2015diffusion}. At odds with many of the influence maximization problems considered in the literature on both static and temporal networks, they show that their optimization problem does not necessarily involve the maximization of a submodular function. They further demonstrate that delaying the activation of some initial spreaders may increase their effective influence.

In this paper, we introduce a discrete-time version of the susceptible-infected-recovered (SIR) model on temporal networks~\cite{pastor2015epidemic}.  We systematically study the influence maximization problem associated with SIR spreading on 12 real-world temporal networks. We test the performance of different approximate algorithms aimed at the identification of the best spreaders in the network. Approximations rely on different levels of dynamical and topological information. Results of our analysis show that having knowledge, even if partial, of the early stages of the network dynamics is essential for an effective prediction of the influential spreaders in temporal networks.

We stress that our modeling framework is very similar to the one previously used by Valdano {\it et al.} for SIS spreading on temporal networks~\cite{valdano2015analytical}. The properties of the two spreading models are however rather different, especially because the SIR model displays a sensitivity to the temporal ordering of the network edges that is stronger than the one observed for the SIS model. Further, both the SI and IC models, previously studied by Habiba and Berger-Wolf~\cite{habiba2007maximizing} and Gayraud \emph{et al.}~\cite{gayraud2015diffusion}, respectively, can be seen as two extreme cases of our model. We extend and generalize those analyses in several respects. First, being able to tune our model between the two extremes, we are essentially able to change the effective level of submodularity of the function that we want to optimize. Second, we do not focus on specific values of the spreading probability for every network. Rather, we tune each network close to its own critical point and study the influence maximization problem near criticality. While studying the phase diagram of the SIR model, we show that the system behavior can be reasonably well predicted by a mean-field approximation and that such an approximation can be effectively used to solve the influence maximization problem. Finally,  we do not assume that the optimization problem is solved using full information about the system. Rather, we systematically test the performance of approximations obtained under partial knowledge of the network topology and dynamics~\cite{erkol2018influence}. Our results, obtained on a corpus of real-word temporal networks that is significantly larger than those typically considered in previous studies, provide clear indications on the type of ingredients that one needs to rely on when available topological information is incomplete or noisy.

%%%%%%%%%%%%%%%%%%%%%%%%%%%%%%%%%%%%%%%%%%%%%%%%%%%%%%%%%%%%%%%%%%%%%%%%%%%%%%%%%%%%%%%%%%%%%%%%%%%%%%%%%%%%%%%%%%%%%%%%%%%%%%%%%%%%%%%%%%%%%%%%%%%%%%%%%%%%%%%%%%%%%%%%

\section{Empirical data and their representation as temporal networks}
%In this study, we focus on temporal networks. 
We focus our attention on 12 empirical datasets containing time-stamped social interactions among pairs of individuals. Datasets refer to two types of interactions. In some of the datasets, interactions correspond to physical proximity contacts, e.g., among high school students~\cite{gemmetto2014mitigation, stehle2011high}, conference attendees~\cite{isella2011s}, and hospital staff/patients~\cite{vanhems2013estimating}. 
In other datasets, interactions stand for emails exchanged by coworkers~\cite{paranjape2017motifs}. In all cases, we treat contacts as undirected. All datasets considered in the study are listed in Table \ref{table:networks}.

Given a dataset, we follow a quite common modeling scheme~\cite{mucha2010community, holme2012temporal, kivela2014multilayer}. We slice the dataset into time windows of identical length $W$. We aggregate all interactions within a time slice to form a temporal network layer. Multiple interactions, between a pair of nodes in the same slice, are reduced to a single unweighted link. All layers contain exactly $N$ nodes, where $N$ is the number of distinct individuals involved in at least one social interaction in the dataset. 
By construction, some nodes may have degree equal to zero in one or more temporal layers. To avoid for the presence of  layers that are too sparsely connected, we exclude from our analysis network layers containing a number of null-degree nodes greater than $0.9\, N$~\footnote{In some datasets, a large portion of layers, corresponding to periods of inactivity, is excluded from the analysis \SM. For example, in the high school temporal data, there are no recorded night-time interactions, and the layers corresponding to such time frames are excluded as they do not contain edges.}. After the cleaning procedure is performed,  the dataset is transformed into $T$ temporal network layers. The $T$ layers are chronologically ordered  from $1$ to $T$. 

We choose different $W$ values depending on the dataset on purpose. Our goal is simply ending up with a similar number $T$ of layers across datasets. Also, the threshold value used for disregarding low-density layers is arbitrarily chosen. We are aware that both ingredients affect the construction of the network layers and the outcome of the spreading process taking place on them. We stress, however, that the goal of the paper is not understanding the dynamics of a specific temporal network and/or a specific choice of the parameters of the spreading model. Rather, we want to study the general problem of influence maximization on temporal networks, and compare different strategies to solve the problem. As far as we are concerned, the real-world datasets considered here just provide useful data for the construction of temporal network topologies, and influence maximization strategies are compared one against the other on the same test sets.

\begin{table}[!htb]
\begin{center}
\begin{tabular}{|l|r|r|r|c|}\hline 
Dataset & $W$ & $T$ & $N$ & Ref. \\\hline 
Email, dept. 1 & $2,880,000$ & $18$ & $309$ & \cite{paranjape2017motifs} \\\hline
Email, dept. 2 & $2,880,000$ & $18$ & $162$ & \cite{paranjape2017motifs} \\\hline
Email, dept. 3 & $2,880,000$ & $18$ & $89$ & \cite{paranjape2017motifs} \\\hline
Email, dept. 4 & $2,880,000$ & $18$ & $142$ & \cite{paranjape2017motifs} \\\hline
High school, 2011 & $14,400$ & $11$ & $126$ & \cite{fournet2014contact} \\\hline
High school, 2012 & $14,400$ & $21$ & $180$ & \cite{fournet2014contact} \\\hline
High school, 2013 & $14,400$ & $14$ & $327$ & \cite{mastrandrea2015contact} \\\hline
Hospital ward & $14,400$ & $20$ & $75$ & \cite{vanhems2013estimating} \\\hline
Hypertext, 2009 & $14,400$ & $11$ & $113$ & \cite{isella2011s} \\\hline
Primary school & $7,200$ & $11$ & $242$ & \cite{gemmetto2014mitigation, stehle2011high} \\\hline
Workplace & $28,800$ & $20$ & $92$ & \cite{genois2015data} \\\hline
Workplace-2 & $28,800$ & $20$ & $217$ & \cite{genois2018can} \\\hline
\end{tabular}
\end{center}
\caption{{\bf Real-world temporal networks.} List of the empirical datasets used to construct temporal networks. From left to right, we report: the name of the dataset, the length $W$ of the temporal window used to slice the data (time is expressed in seconds), the number $T$ of network layers resulting after slicing and cleaning data, the number of nodes $N$ in the network, and the reference to the paper(s) where the data were first considered.}
\label{table:networks}
\end{table}

At the end of the above-described procedure, we have at our disposal a sequence %$\vec{A} = 
$\{ A^{(1)}, \ldots, A^{(t)}, \ldots, A^{(T)} \}$
of $T$ temporal adjacency matrices. The adjacency matrix $A^{(t)}$ fully encodes information about the topology of the $t$th temporal network layer, with its generic element $A^{(t)}_{ij} = A^{(t)}_{ji}  = 1$ if a connection exists between nodes $i$ and $j$ at stage $t$ of the network dynamics, whereas $A^{(t)}_{ij} = A^{(t)}_{ji}  = 0$, otherwise. 
%In our analysis, we adopt two main ways of neglecting temporal topological information. First, we randomize the order of the layers. Second, we will consider a flat version of the network, where a single static network is obtained by aggregating all layers. 

\section{Spreading dynamics}

We study spreading dynamics taking place on temporal networks. In analogy with the work by Valdano {\it et al.}~\cite{valdano2015analytical},
we assume that the characteristic timescales of the spreading process and of the network evolution are identical. We consider the discrete-time version of the SIR model to mimic spreading dynamics \cite{pastor2015epidemic}, thus differentiating from the work by Valdano {\it et al.} where the SIS model was considered instead. In the SIR model, the state  $\sigma^{(t)}_i$ for node $i$ at time $t$ can be $\sigma^{(t)}_i =S, I, $ or $R$. Every  node $i$ that is infected at time $t$, i.e., $\sigma^{(t)}_i = I$, attempts to infect all its susceptible neighbors, i.e., all $j$ such that $A^{(t)}_{ij} = 1$ and $\sigma^{(t)}_j = S$. Infection is successfully transmitted with probability $\lambda$. In case of a successful attempt, the state of the newly infected node $j$ changes as $\sigma^{(t)}_j = S \to \sigma^{(t+1)}_j = I$, meaning that the node can spread the infection from time $t+1$ on. After all spreading attempts have been performed, each infected node $i$ may recover with probability $\mu$. A successful recovery attempt changes the state of node $i$ as $\sigma^{(t)}_i = I \to \sigma^{(t+1)}_i = R$. A recovered node does no longer participate in the dynamics, thus it cannot spread nor receive the infection. After all recovery attempts have been performed, time increases as $t \to t+1$. 
%The process starts from arbitrary initial condition, and is repeated for a certain number of iterations or till the system reaches a steady-state configuration.

Two standard models of spreading are obtained as special cases of our more general model. If $\mu =1$, the model reduces to the IC model~\cite{gayraud2015diffusion}. If $\mu =0$ and no nodes are initially in the recovered state, then the SIR model reduces to the discrete-time version of the SI model~\cite{habiba2007maximizing}.

\begin{figure}[!htb]
\centering
\includegraphics[width=0.45\textwidth]{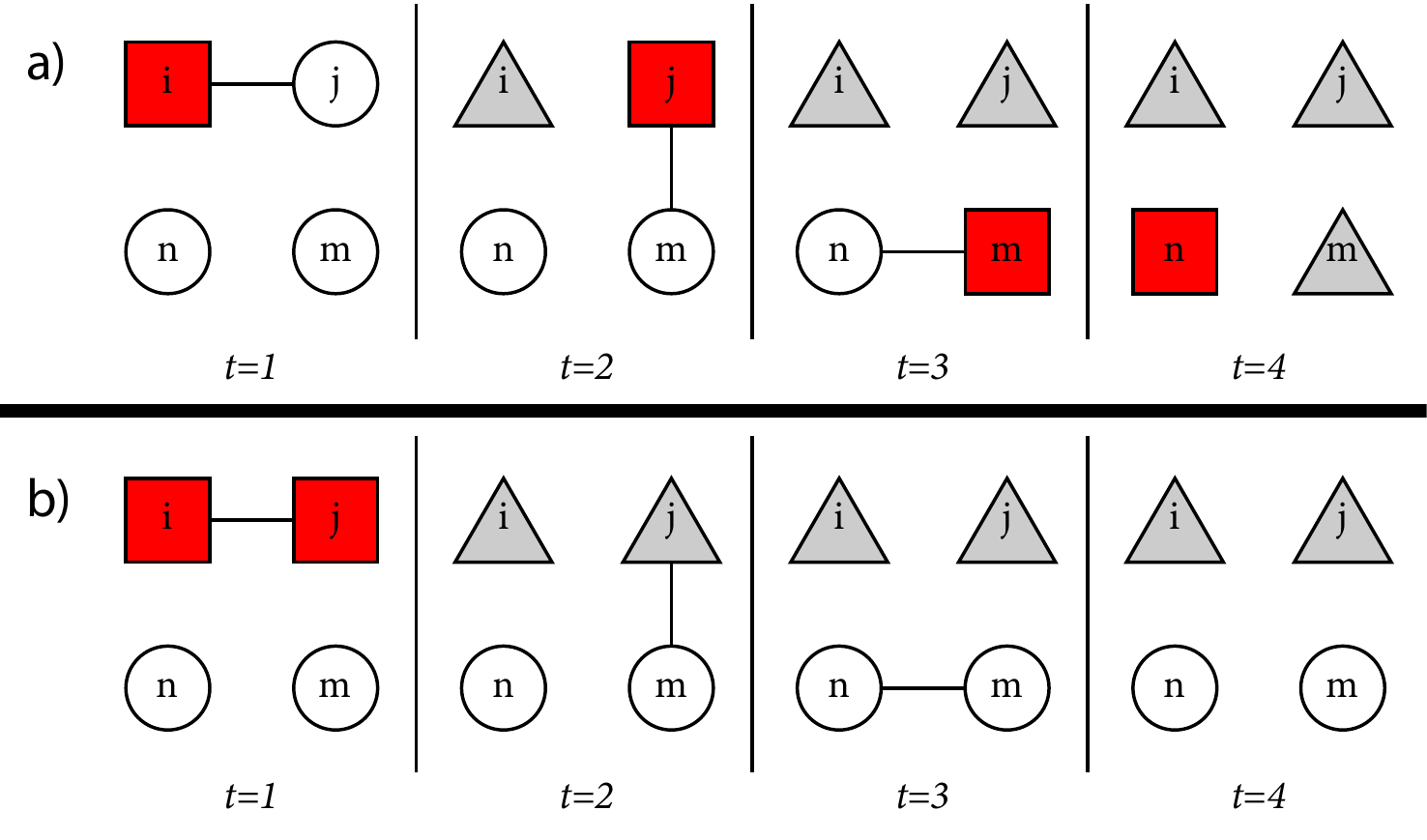}

\caption{{\bf Susceptible-infected-recovered model on temporal networks.} Illustrative example of the modeling framework proposed in this paper, where SIR spreading occurs on a temporal network. In the example, the network consists of four nodes and four temporal layers, and the spreading dynamics takes place over four discrete temporal stages. For simplicity, in the illustration we set 
the SIR model parameters $\lambda = \mu = 1$ so that the dynamics is  deterministic. (a) The initial condition is such that only node $i$ is infected, while all others are in the susceptible state. At the end of the dynamics, all nodes are either infected or recovered. (b) Nodes $i$ and $j$ are initially infected, and they are recovered in the final configuration. Nodes $n$ and $m$ remain in the susceptible state.}
\label{fig:1}
\end{figure}

%\section{Sensitivity to initial conditions}
Starting from a given initial configuration $\vec{\sigma}^{(1)} = (\sigma_1^{(1)}, \ldots, \sigma_N^{(1)} )^\intercal$, we follow the dynamics of the model until the last iteration of spreading is performed, reaching the final configuration $\vec{\sigma}^{(T+1)}$. Even for fixed values of $\lambda$ and $\mu$, the outcome of the spreading model is highly sensitive to the initial conditions (see Fig.~\ref{fig:1}). 

We restrict our attention to special types of initial configurations where all nodes are in the $S$ state, with the exception of a set $\mathcal{X} = \{x_1, x_2, \ldots, x_{|\mathcal{X}|}\}$ of seed nodes that are in the $I$ state, i.e., $\sigma^{(1)}_{i}=I$ if $i \in \mathcal{X}$ and $\sigma^{(1)}_{i}=S$ if $i \notin \mathcal{X}$. Given the parameters of the SIR model and the topology of the temporal network, we estimate the relative outbreak size $O(\mathcal{X})$ generated by the seed set $\mathcal{X}$ in a single realization of the SIR model. $O(\mathcal{X})$ is defined as the total number of nodes found either in the $R$ or $I$ state in the stage $T+1$ of the process, divided by the network size $N$, i.e.,
\begin{equation}
O (\mathcal{X}) = \frac{1}{N} \; \sum_{i =1}^{N} \, \left( \mathds{1}_{\sigma_i^{(T+1)}, I} + \mathds{1}_{\sigma_i^{(T+1)}, R} \right) \; ,
\label{eq:outbreak}
\end{equation}
with $\mathds{1}_{x,y}$ identity operator, i.e., $\mathds{1}_{x,y} = 1$ if $x=y$ and $\mathds{1}_{x,y} = 0$, otherwise. $O (\mathcal{X})$ is a random variable, obeying some probability distribution. We stress that $O (\mathcal{X})$ strongly depends on the choice of the parameters $\lambda$ and $\mu$, the topology of the network layers and their time ordering. However, we do not report the explicit dependence on these factors for shortness of notation. 

As in many of the papers on influence maximization~\cite{kempe2003maximizing, erkol2019systematic}, we use the average value of the outbreak size as the metric of influence for the seed set $\mathcal{X}$. Specifically, we numerically estimate the influence of the seed set $\mathcal{X}$ over a finite number $Q$ of numerical simulations as
\begin{equation}
\left\langle O (\mathcal{X}) \right\rangle = \frac{1}{Q} \sum_{q=1}^Q \; O_q (\mathcal{X}) \; ,
\label{eq:influence}
\end{equation}
where $O_q (\mathcal{X})$ is the relative outbreak size of Eq.~(\ref{eq:outbreak}) obtained in the $q$th instance of the model. We use $Q = 2,000$ in all our numerical results, unless otherwise specified.

\begin{figure}[!htb]
\centering
\includegraphics[width=0.45\textwidth]{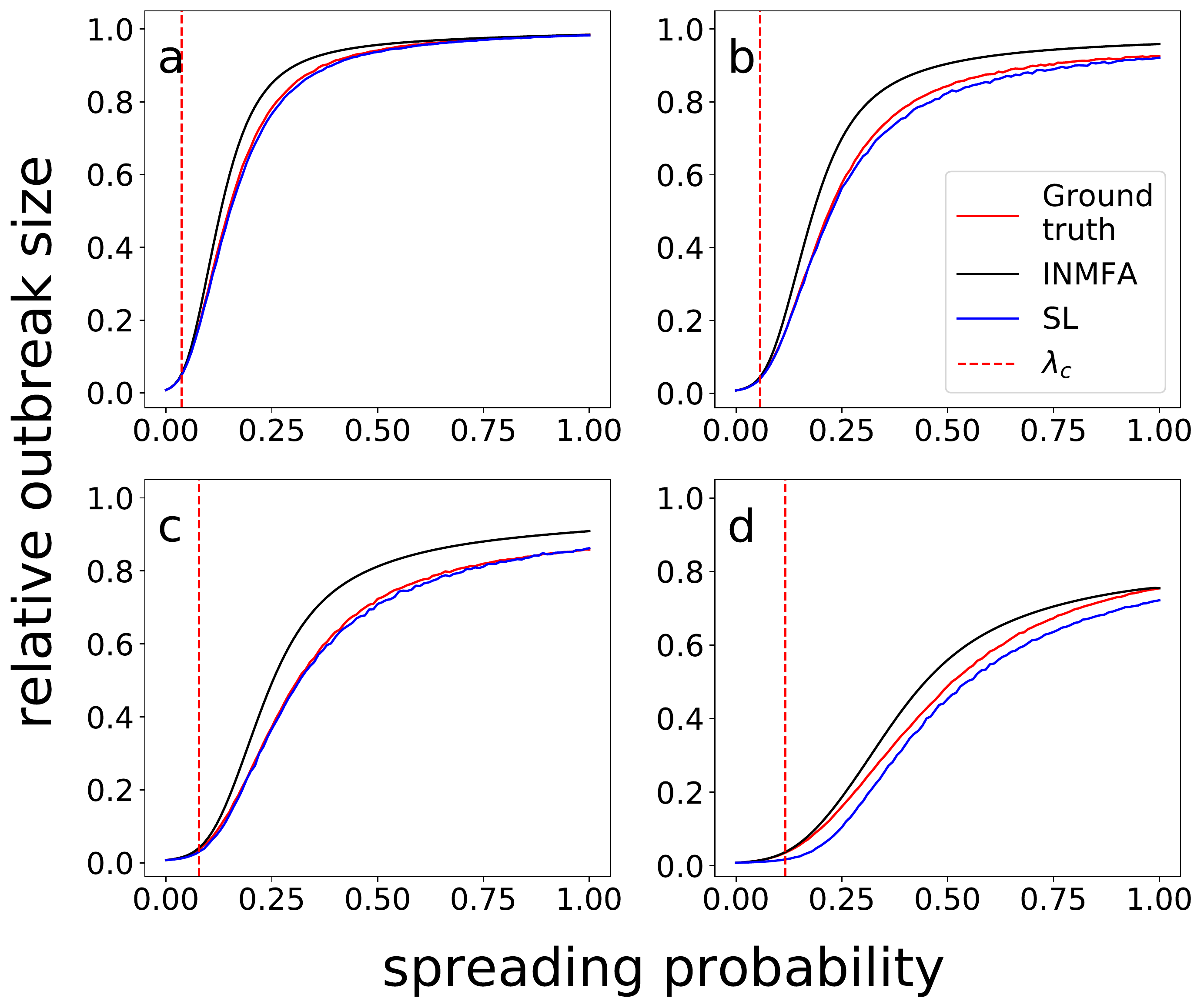}
\caption{{\bf Epidemic transition in real-world temporal networks.} (a) Average value of the relative outbreak size $\left\langle O (\mathcal{X}) \right\rangle$ as a function of the spreading probability $\lambda$. The seed set corresponds to one randomly chosen node. Results are obtained on the "High school, 2011" network, and by setting $\mu =0$. Results from numerical simulations on the real network topology (red curve) are compared against those predicted by INFMA (black curve). The dashed red line indicates the position of our best estimate of the critical value of the spreading probability, i.e., $\lambda_c$. We further display results of numerical simulations obtained on the same network topology but with the order of the temporal network layers randomized (SL, blue curve).  (b) Same as in (a), but for $\mu = 0.25$. (c) Same as in (a), but for $\mu = 0.5$. (d) Same as in (a), but for $\mu = 1$.}
\label{fig:2}
\end{figure}

In Fig.~\ref{fig:2}, we show typical phase diagrams obtained for seed sets of size one. In the diagrams, a data point of the outbreak size for a given pair of parameter values $\mu$ and  $\lambda$ is obtained as follows. We consider $N$ different initial conditions, each corresponding to one of the nodes selected as the initial spreader with all other nodes initially in the susceptible state, i.e., $\mathcal{X} = \{i\}$ for all $i = 1 , \ldots, N$. For each initial condition, we run $Q=500$ simulations and estimate the influence of the node according to Eq.~(\ref{eq:influence}). We finally take the average value of the influence over all initial conditions as representative quantity for the system outbreak size. The system is characterized by a phase transition from a nonendemic to an endemic phase as the parameters of the spreading model are varied. Specifically, the endemic phase is obtained  for sufficiently large values of the spreading probability $\lambda$. The critical $\lambda$ value, namely $\lambda_c$, where the transition occurs is a function of the recovery probability $\mu$, i.e., $\lambda_c = \lambda_c(\mu)$. We note that $\lambda_c$ increases as $\mu$ increases. We stress that the system size is finite here, so we are not facing a genuine phase transition. Nonetheless, the change in the value of average outbreak size lets us easily notice the presence of a regime where the outbreak is confined to a small part of the network and a regime where spreading involves a large portion of the system.

We estimate the critical value of the spreading probability $\lambda_c(\mu)$ for a given value of the recovery probability $\mu$ as the $\lambda$ value that maximizes the ratio between standard deviation and average value of the outbreak size, both computed over $Q = 500$ numerical simulations of the spreading process initiated by a single randomly chosen seed (we use the same procedure as described above, but for simplicity, only nodes with at least one connection in the first layer of the network are considered as possible seeds). We note that looking at the peak of the ratio standard deviation over average value is not the only possible way of defining and identifying the critical point of the transition. One, for example, may  look at the position of the peak of the standard deviation only. In general, different definitions may lead to slightly different estimates of $\lambda_c$. We stress, however, that the  $\lambda_c$ values we obtain seem to identify quite accurately the transition point (see Fig.~\ref{fig:2}) and that the exact value of the transition point is not very crucial for the type of analysis we are performing in this paper. In Table~\ref{table:networks_critical}, we report the $\lambda_c$ values obtained for the various networks considered in this paper.  

\begin{table}[!htb]
\begin{center}
\begin{tabular}{|l|r|r|r|r|r|}\hline 
Network & $\lambda_c (\mu = 0)$ & $\lambda_c (\mu = 0.25)$ &$\lambda_c (\mu = 0.5)$ & $\lambda_c (\mu = 1)$ \\\hline 
Email, dept. 1 & 0.016 & 0.043 & 0.069 & 0.130 \\\hline
Email, dept. 2 & 0.010 & 0.027 & 0.049 & 0.096 \\\hline
Email, dept. 3 & 0.016 & 0.038 & 0.066 & 0.123 \\\hline
Email, dept. 4 & 0.010 & 0.029 & 0.047 & 0.099 \\\hline
High school, 2011 & 0.037 & 0.057 & 0.078 & 0.116 \\\hline
High school, 2012 & 0.025 & 0.077 & 0.136 & 0.205 \\\hline
High school, 2013 & 0.023 & 0.042 & 0.064 & 0.119 \\\hline
Hospital ward & 0.017 & 0.048 & 0.087 & 0.207 \\\hline
Hypertext, 2009 & 0.023 & 0.041 & 0.060 & 0.097 \\\hline
Primary school & 0.013 & 0.019 & 0.029 & 0.043 \\\hline
Workplace & 0.042 & 0.123 & 0.241 & 0.308 \\\hline
Workplace-2 & 0.023 & 0.063 & 0.119 & 0.248 \\\hline
\end{tabular}
\end{center}
\caption{{\bf Critical thresholds of real-world temporal networks.} We report our numerical estimates of the critical spreading probability $\lambda_c(\mu)$ for the temporal networks of Table~\ref{table:networks}. Different columns correspond to different values of the recovery probability $\mu$. Errors associated to the estimates are all equal to or smaller than $10^{-3}$ and they are not reported in the table for sake of compactness.}
\label{table:networks_critical}
\end{table}

\begin{figure}[!htb]
\centering
\includegraphics[width=0.45\textwidth]{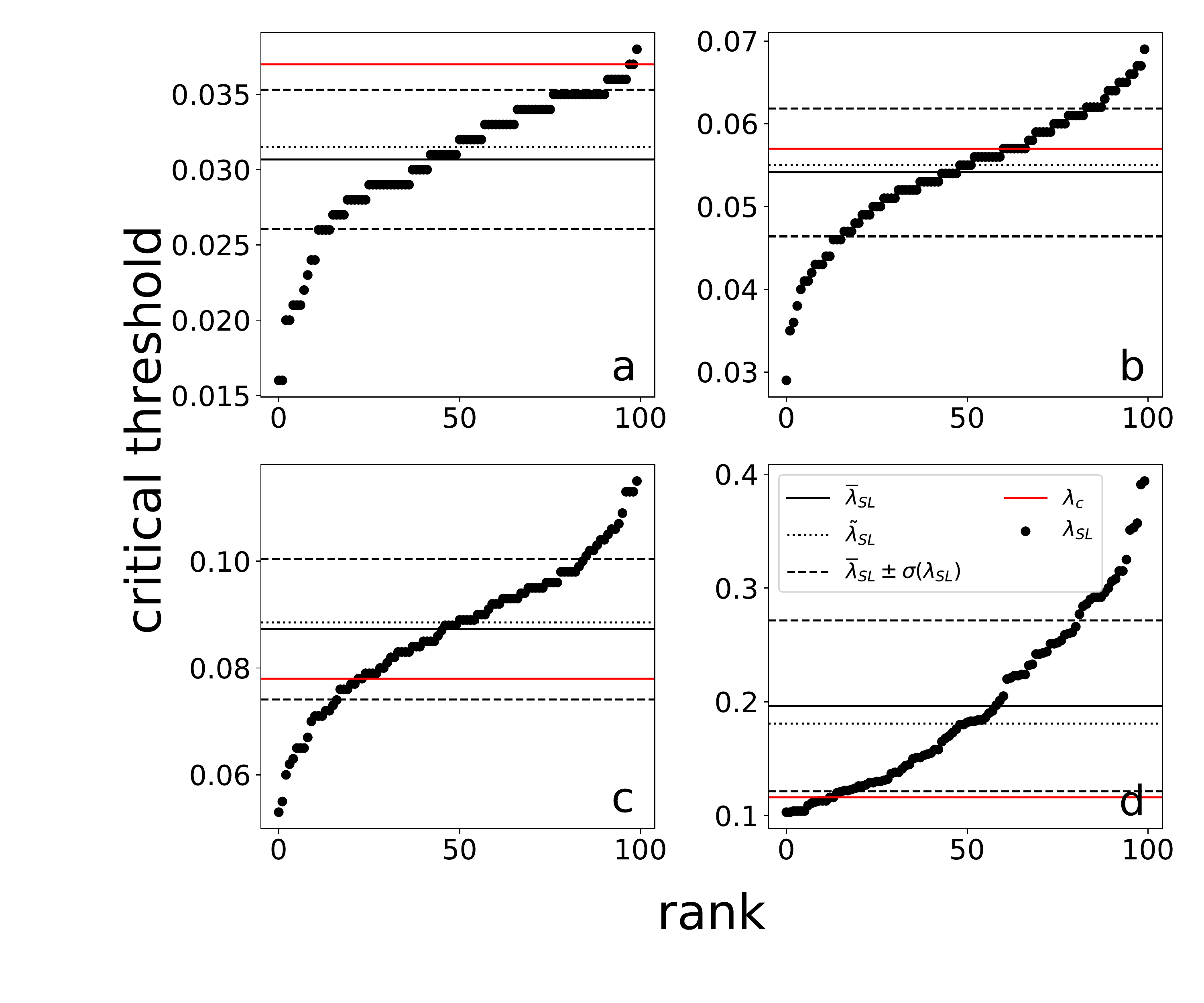}
\caption{{\bf Sensitivity of the spreading outcome to network dynamics.} (a) Best estimates of the critical spreading probability $\lambda_{SL}$ for randomized versions of the "High school, 2011" temporal network. SIR recovery probability is $\mu = 0$. The randomization consists in reordering the temporal layers only, while the topology of the individual layers is kept invariant. Each black circle corresponds to a specific realization of the randomization process. In the visualization, we simply sort the various realizations depending on their $\lambda_{SL}$ value. We display horizontal lines identifying the average $\bar{\lambda}_{SL}$ (full black line), the region corresponding to one standard deviation away from the mean [$\bar{\lambda}_{SL} \pm \sigma(\lambda_{SL})$, dashed black lines], the median value $\tilde{\lambda}_{SL}$ (dotted black line), and the actual critical value $\lambda_c$ measured on the nonrandomized version of the network (red full line, Table~\ref{table:networks_critical}). (b) Same as in (a), but for $\mu = 0.25$. (c) Same as in (a), but for $\mu = 0.5$. (d) Same as in (a), but for $\mu = 1$.}
\label{fig:3}
\end{figure}

We note that systems display sensitivity to the temporal organization of the underlying networks, and the sensitivity is more apparent for large $\mu$ values than for small $\mu$ values. Phase diagrams, and resulting values of the spreading probability where transitions occur, may  dramatically vary by simply randomizing the order of the layers but without changing the actual network topology of the layers (see Fig.s~\ref{fig:2} and \ref{fig:3} and \SM). This is a quite remarkable difference with respect to the SIS modeling framework by Valdano {\it et al.}, where the actual order of the temporal layers is not as important for the outcome of the spreading dynamics~\cite{valdano2015analytical}.

\section{Individual-node mean-field approximation}

We can provide a relatively simple description of the spreading process using the 
individual-node mean-field approximation (INMFA)~\cite{pastor2015epidemic}. The approximation consists in describing the stochastic state variable $\sigma_i^{(t)}$ for node $i$ at time $t$ with the three deterministic variables $S_i^{(t)}$, $I_i^{(t)}$  and $R_i^{(t)}$. Specifically, we have that $S_i^{(t)} = \textrm{Prob.}[\sigma_i^{(t)} = S] $, i.e., the probability to find node $i$ in state S at stage $t$ over an infinite number of realizations of the process. Similarly,  we have that $I_i^{(t)} = \textrm{Prob.}[\sigma_i^{(t)} = I] $ and $R_i^{(t)} = \textrm{Prob.}[\sigma_i^{(t)} = R] $. The three deterministic variables are related by the constraint $I_i^{(t)} + S_i^{(t)} + R_i^{(t)} = 1$. Further, the approximation consists in neglecting dynamical correlations among two or more state variables, so that joint probabilities can be replaced by products among marginal probabilities. For example, the probability at stage $t$ of the dynamics to find nodes $i$ and $j$ respectively in the infected and recovered states is simply written as $\textrm{Prob.}[\sigma_i^{(t)} = I, \sigma_j^{(t)} = R] = I_{i}^{(t)} \, R_{j}^{(t)}$.

Under INMFA, we can describe SIR dynamics on the temporal network with the following set of coupled equations
\begin{equation}
\begin{array}{ll}
I_i^{(t)} = & (1-\mu)I_i^{(t-1)} + 
\\ & \left( 1-I_i^{(t-1)}-R_i^{(t-1)}\right) \,  \left[ 1-\prod_j \left(1-\lambda A_{ji}^{(t-1)}I_j^{(t-1)}\right)\right] 
\end{array}
\label{eq:mf1}
\end{equation}
and
\begin{equation}
R_i^{(t)}=R_i^{(t-1)}+\mu I_i^{(t-1)} \; . \label{eq:mf2}
\end{equation}
The initial conditions are suitably chosen depending on the problem at hand. In our case, we set $I_i^{(1)}=1$ for every $i \in \mathcal{X}$ and $S_i^{(1)}=1$ for $i \notin \mathcal{X}$, and use the above equations to obtain solutions for $t>1$.
Equation~(\ref{eq:mf1}) simply tells us that the probability that node $i$ is in the infected state at time $t$ is the sum of two terms: (i) The probability that the node was already in the infected state at the previous stage of the dynamics but did not recover; and (ii) the probability that the node was not already infected but received the infection by at least one of its infected neighbors at the previous time step. Equation~(\ref{eq:mf2}) instead tells us that the probability that node $i$ is in the recovered state at time $t$ is the sum of the probability that the node was already recovered or just recovered at the previous stage of the dynamics.
INMFA neglects dynamical correlation between nodes.
Variables are treated as dynamically independent when instead they are not. In particular, there is a nonnull probability that spreading may occur simultaneously in opposite directions along the same edge, thus causing a systematic overestimation of the true probability of infection.
Starting from the imposed initial conditions, 
one iterates Eqs.~(\ref{eq:mf1}) and~(\ref{eq:mf2}) to obtain the marginal probabilities of all nodes in the network at a given stage of the dynamics. The relative size of the outbreak at time $t$ is obtained by simply taking the sum
\begin{equation}
O_{INMFA}^{(t)} = \frac{1}{N} \, \sum_{i=1}^{N} \,
\left( I_i^{(t)} + R_{i}^{(t)} \right) \; .
\label{eq:mf3}
\end{equation}

In Fig.\ref{fig:2}, we compare results from the INMFA with ground-truth values obtained from numerical simulations of the spreading process. 
Due to the independence among variables that is assumed in INMFA, the approximation
overestimates the true outbreak size, and under INMFA the phase transition is expected to happen earlier than in the true dynamical system. 
Nonetheless, we note that INMFA provides a relatively good prediction of the true system outcome, especially in the subcritical regime and around criticality.

\section{Influence maximization}

We consider the classical problem of influence maximization consisting in finding the set of seed nodes that maximizes the average size of the outbreak~\cite{domingos2001mining, kempe2003maximizing}. The maximization problem is solved with a constraint on the size of the seed set, and for a given choice of the spreading parameters $\lambda$ and $\mu$. 

\subsection{Greedy optimization}

As influence maximization is a NP-hard problem~\cite{kempe2003maximizing}, its solutions can only be approximated.  On static networks, the best available strategy to solve the problem of influence maximization consists of a greedy algorithm~\cite{kempe2003maximizing, erkol2019systematic}. 
The mechanism of the algorithm is quite simple and naturally extends to temporal networks~\cite{gayraud2015diffusion}. Indicated with $\mathcal{B}_v = \{b_1, b_2, \ldots, b_v\}$ the seed set identified by the algorithm at stage $v$, we initialize the algorithm with $\mathcal{B}_0 = \emptyset$. Then, for $v > 0$ we have 
\begin{equation}
b_v = \arg \; \max_{x \notin \mathcal{B}_{v-1}} \; \left\langle O (\mathcal{B}_{v-1} \cup \{x\}) \right\rangle \; .
\label{eq:greedy_algorithm}
\end{equation}
Essentially, the best seed set is built sequentially by adding one node at a time. The node $b_v$ selected at stage $v$ is the one providing the largest marginal increment of influence to the existing seed set. We stress that, at each stage $v$ of the algorithm, one needs to numerically estimate $\left\langle O (\mathcal{B}_{v-1} \cup \{x\}) \right\rangle$ 
for all $x \notin \mathcal{B}_{v-1}$ in order to select $b_v$ appropriately, and
simulations must be run independently for each potential seed set $\mathcal{B}_{v-1} \cup \{x\}$. We remark that the method requires as inputs full topological and dynamical information about the system, including the actual values of the parameters of the spreading model. In the following, we denote solutions of the influence maximization problem obtained via the standard greedy optimization method with the acronym GR.

For the SIR model in static networks, it is known that influence is a growing and submodular function~\cite{kempe2003maximizing}, thus greedy solutions are guaranteed to be within a margin $1-1/e \simeq 0.63$ from the true optimum~\cite{nemhauser1978analysis}. 
On temporal networks, the two above conditions are valid only for the special case $\mu = 0$~\cite{habiba2007maximizing}. However for $\mu >0$, influence may decrease as the system size increases, so that the function may also violate the submodularity property (see Ref.~\cite{gayraud2015diffusion} and Fig.~\ref{fig:1} for a specific example). As a consequence, the greedy algorithm doesn't guarantee a known optimality gap. 
 
 \subsection{Approximate greedy optimization}
 
 The complexity of the algorithm described in Eq.~(\ref{eq:greedy_algorithm}) grows cubically with the network size, as estimates of the influence of all seed sets are obtained via numerical simulations. Complexity reduction is possible by approximating  $\langle O (\mathcal{X}) \rangle$ in some way, so that the elementary choice of Eq.~(\ref{eq:greedy_algorithm}) is replaced by
 \begin{equation}
b_v = \arg \; \max_{x \notin \mathcal{B}_{v-1}} \; F(\mathcal{B}_{v-1} \cup \{x\}) \; .
\label{eq:greedy_algorithm_approx} 
\end{equation}
Here we indicated with $F(\mathcal{B} \cup \{x\})$ a generic function that estimates the incremental importance of node $x \notin \mathcal{B}$ for the influence of the set $\mathcal{B}$, assuming that influence is not directly measured. Typical choices of $F$ leverage  parallel and/or partial computation to decrease computational complexity. For the IC model on static networks for example, the equivalence of the spreading model with static bond percolation suggests how to decrease algorithmic complexity without sacrificing performance \cite{chen2009efficient, hu2018local}. In Ref.~\cite{chen2009efficient}, 
$F(\mathcal{B} \cup \{x\})$ is defined as the average size of the clusters that contain node $x$ but no nodes already in $\mathcal{B}$, a quantity that is equivalent to the targeted ground truth $\left\langle O (\mathcal{B} \cup \{x\}) \right\rangle$ and that can be computed in parallel for all nodes. Variations of the methods of Refs.~\cite{chen2009efficient, hu2018local} are not easily implementable for the general SIR model, and the temporal nature of the network creates additional challenges. We implement, however, an approximate version of greedy optimization that uses the INMFA prediction of Eq.~(\ref{eq:mf3}) for the definition of the function $F$. Many other methods aiming at reducing algorithmic complexity use network centrality metrics for the definition of $F$ such that, during the course of algorithm, the score is static (e.g., degree centrality) or can be quickly recomputed with partial computation (e.g., adaptive degree centrality). On the basis of  previous analyses conducted on static networks~\cite{erkol2019systematic}, we focus our attention on adaptive degree centrality only.

\subsection{Greedy optimization under incomplete information}

In the sections above, we made the strong hypothesis that optimization is performed by knowing in advance that the network is evolving, and how it exactly evolves. Further, the optimization is performed by being aware of the true spreading dynamics, including the actual values of the model parameters and the existence of a specific temporal horizon in the spreading process.

Having full knowledge of all the ingredients of the problem is, however, a strong assumption. In realistic scenarios, it is much more likely to attempt to solve the problem with limited and/or noisy information. 
For example, we may have at our disposal only a flat and aggregated version of the true network, where temporal information is absent. In this scenario, we would apply the greedy algorithm to a static network, disregarding completely the existence of network evolution and the time horizon for the spreading. We would further be able to identify the critical regime of spreading for the static network only. In essence, we would use still the same approach as Eq.~(\ref{eq:greedy_algorithm_approx}) where the function $F$ represents an approximation of the ground-truth $\langle O \rangle$ of Eq.~(\ref{eq:greedy_algorithm}). However, the approximation would not be made with the goal of reducing computational complexity. Rather, it would be enforced by the incompleteness of the information at our disposal in the solution of the true problem. 

There are many potential ways in which topological information may arrive to us incomplete or noisy. We consider several possibilities listed in  Table~\ref{table:methods}. The simplest setting is what we call SL, where layers are randomly reordered, but all other information required for the solution of the influence maximization problem is preserved. SL is the same setting already considered in the study of the sensitivity of the outbreak size to the temporal ordering of the network layers  (Fig.s~\ref{fig:2} and~\ref{fig:3}). We remark that, in the SL setting, we still rely on the same exact scheme as described for greedy optimization. Thus, to perform the selection step of Eq.~(\ref{eq:greedy_algorithm_approx}), we run multiple numerical simulations of SIR dynamics by assuming that we know the true values of the spreading and recovery probabilities of the spreading model, but also that we believe that the true network dynamics is given by the specific SL setting at our hand. 

We then consider cases where part of the temporal information is not present in our input data. For example, we consider the setting FL  where only the first temporal layer is used in the solution of the problem. The setting RL is analogous to FL with the only difference that one randomly chosen layer is selected to play the role of the static network. In these cases, we  perform standard greedy optimization under the hypothesis of having a static network topology, and we assume that the critical value of the spreading probability is given by the one of the static network. 

We then consider a scenario where temporal information is flattened. The solution to the influence maximization problem relies on an aggregated version of the network and no temporal horizon is provided in the estimate of the outbreak size. We name this setting ST. We  perform standard greedy optimization under the hypothesis of having a static network topology, and we assume that the critical value of the spreading probability is the one valid for the aggregated static network. 

Also, we consider further approximations where the optimization problem is solved using adaptive degree (AD) centrality computed using full or partial  information of the system topology. AD is a metric similar to degree centrality. The only difference is that, once a node is selected as a seed, the node is considered as removed from the network and the degree values of all other nodes are recomputed before selecting the next seed. Essentially, the function $F(\mathcal{B}_{v-1} \cup \{x\} )$ appearing in Eq.~(\ref{eq:greedy_algorithm_approx}) equals the number of connections of node $x$ with nodes that do not belong to the set $\mathcal{B}_{v-1}$. Specifically, we consider the approximations AD-F and AD-A where respectively the first layer or the aggregation of all layers are used for the computation of the adaptive degree centrality. The method in this case relies on topological information only, and there is no need of feeding the algorithm with information about the spreading model and its parameter values.

Finally, we consider the setting RND, where the seed set is built by randomly selecting nodes. This is the only viable option in case nothing is known about the network, and should provide the worst performance possible.

\begin{table}[!htb]
\begin{center}
\begin{tabular}{|l|l|l|l|}\hline
Approximation & Time horizon & Time order & Temporal layers %& $\lambda_c$ 
\\\hline 
GR & yes & yes & all %& $\lambda_c(\mu)$ 
\\\hline
INMFA & yes & yes & all %& $\lambda_c(\mu)$ 
\\\hline
RND & no & no & none %& - 
\\\hline 
SL & yes & no & all %& $\lambda_c(\mu)$ 
\\\hline
FL & no & no & one %& $\lambda_c^{(t=1)}(\mu)$ 
\\\hline
RL & no & no & one %& $\lambda_c^{(t=i)}(\mu)$ 
\\\hline
ST & no & no & aggregate %& $\lambda_c^{agg}(\mu)$ 
\\\hline
AD-F & no & no & one %& - 
\\\hline
AD-A & no & no & aggregate %& - 
\\\hline
\end{tabular}
\end{center}
\caption{{\bf Identification of influential spreaders in temporal networks.} We list here the various approximations used in the solution of the influence maximization problem. From left to right, the columns of the table report: the acronym of the approximation, awareness by the approximation about the existence of a temporal horizon in the spreading, awareness by the approximation about the temporal evolution of the network topology, number/type of temporal layers used in the approximation.
%, $\lambda$ values used in identifying influential spreaders. 
The various approximations are described in the main text.} 
\label{table:methods}
\end{table}

\subsection{Systematic tests of performance}

We apply the different approximations of Table~\ref{table:methods} in the identification of the influential spreaders in the 12 temporal networks constructed from the datasets of Table~\ref{table:networks}.
We consider four distinct values of the recovery probability $\mu=0$, $\mu=0.25$, $\mu=0.5$, and $\mu = 1$. For each network and $\mu$ value, we consider the critical value $\lambda_c(\mu)$ of the spreading probability, see Table~\ref{table:networks_critical}. We finally consider separately the cases $\lambda = 0.5 \, \lambda_c (\mu)$, $\lambda = \lambda_c (\mu)$, and $\lambda = 2 \, \lambda_c (\mu)$ as representative for the subcritical, critical, and supercritical dynamical regimes, respectively. In summary, for each of the $12$ temporal networks we consider $12$ distinct combinations of $\mu$ and $\lambda$ values, so that each approximation for the solution of the influence maximization problem is tested in $144$ different experimental settings.

We stress that the true values of the spreading probability are used only for predictions under the GR, INMFA and SL settings. The other methods assume different critical values for $\lambda$, and predictions for the various regimes are made using such a value as a reference. Predictions of all approximations are tested on the ground-truth dynamics. That is, given a set of predicted seeds, we run numerical simulations of the spreading process using true parameter values on the true temporal network.

\begin{figure}[!htb]
\centering
\includegraphics[width=0.45\textwidth]{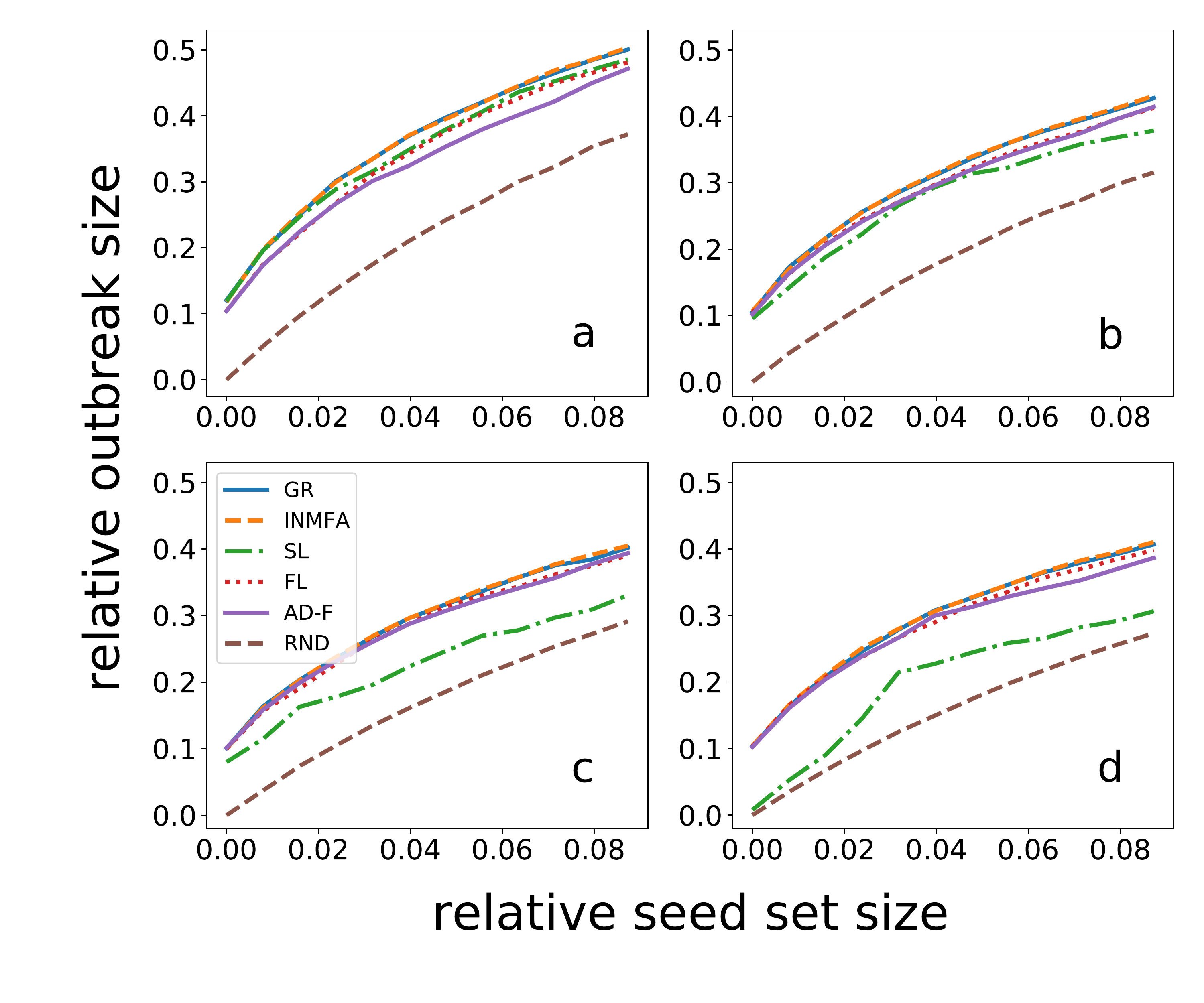}
\caption{{\bf Identification of influential spreaders in temporal networks.} (a) Average value of the relative size of the outbreak, i.e.,  $\left\langle O (\mathcal{X}) \right\rangle$ [see Eq.~(\ref{eq:influence})], as a function of the relative size of the seed set, i.e., $|\mathcal{X}| / N$. The seed set is selected according to some of the approximations described in the text and listed in Table~\ref{table:methods}. The network analyzed is "High school, 2011." Spreading dynamics is critical, with recovery probability $\mu =0$ and $\lambda=\lambda_c(\mu)=0.037$. (b) Same as in panel a, but for $\mu = 0.25$ and $\lambda=\lambda_c(\mu)=0.057$. (c) Same as in panel a, but for $\mu = 0.5$ and $\lambda = \lambda_c(\mu)=0.078$. (d) Same as in panel a, but for $\mu = 1$ and $\lambda=\lambda_c(\mu)=0.116$.}
\label{fig:sample_plot}
\end{figure}

A typical outcome of the systematic analysis we perform is displayed in Fig.~\ref{fig:sample_plot}. There, we plot the average value of the relative outbreak size as a function of the relative size of the seed set.  We display results only for a selection of the approximations listed in Table~\ref{table:methods}, and only for the critical regime of spreading. Results for the other methods, and for other dynamical regimes, are reported in Ref.~\SM. The best solution is generally obtained by GR, i.e., the straight implementation of the greedy optimization strategy of Eq.~(\ref{eq:greedy_algorithm}). This is not surprising as the method relies on complete topological and dynamical information.
Although $\left\langle O \right\rangle$ is not submodular, the shape of the curve indicates an effective submodular behavior resulting from the maximization process. Results obtained under INMFA are generally very close to (sometimes even slightly better than) those of GR. As one may expect, the worst performance is obtained by random selection, i.e., RND. Using FL provides a quite good performance despite other temporal information being neglected. SL, where one is aware of network evolution, but does not know exactly how the temporal layers are ordered, displays poor performance. AD-F provides a fair approximation in terms of performance even though it has no information on spreading dynamics.

We use GR as the baseline for assessing the quality of the solutions obtained under the other approximations~\cite{erkol2018influence, erkol2019systematic}. Given a network with $N$ nodes, we fix the targeted seed set size to $V = 0.1 \, N$. For the generic approximation $a$, we evaluate the area under the curve of the spreading impact of the seed set $\mathcal{B}^{(a)}$ that the approximation identifies, i.e., 
\begin{equation}
\textrm{AUC}_a = \sum_{v=1}^{V} \, \langle O (\mathcal{B}^{(a)}_v ) \rangle \; ,
\label{eq:AUC}
\end{equation}
where $\mathcal{B}^{(a)}_v$ is the seed set found by the approximation $a$ in the $v$th step of the optimization algorithm of Eq.~(\ref{eq:greedy_algorithm_approx}). The definition can be clearly adapted to compute $\textrm{AUC}_\textrm{GR}$, thus leading to a metric of performance for the straight implementation of the greedy algorithm of Eq.~(\ref{eq:greedy_algorithm}). We note that the metric $\textrm{AUC}_{a}$ gives importance to the global impact of the seed set $\mathcal{B}^{(a)}_V$ but also to the order in which nodes are placed in the set during the optimization steps of Eq.~(\ref{eq:greedy_algorithm_approx}). We then normalize the performance $p_a$ of the generic approximation $a$ with GR by simply taking the ratio
\begin{equation}
p_a = \frac{\textrm{AUC}_a}{\textrm{AUC}_{\textrm{GR}} }  \; .
\label{eq:performance}
\end{equation}

\begin{figure}[!htb]
\centering
\includegraphics[width=0.45\textwidth]{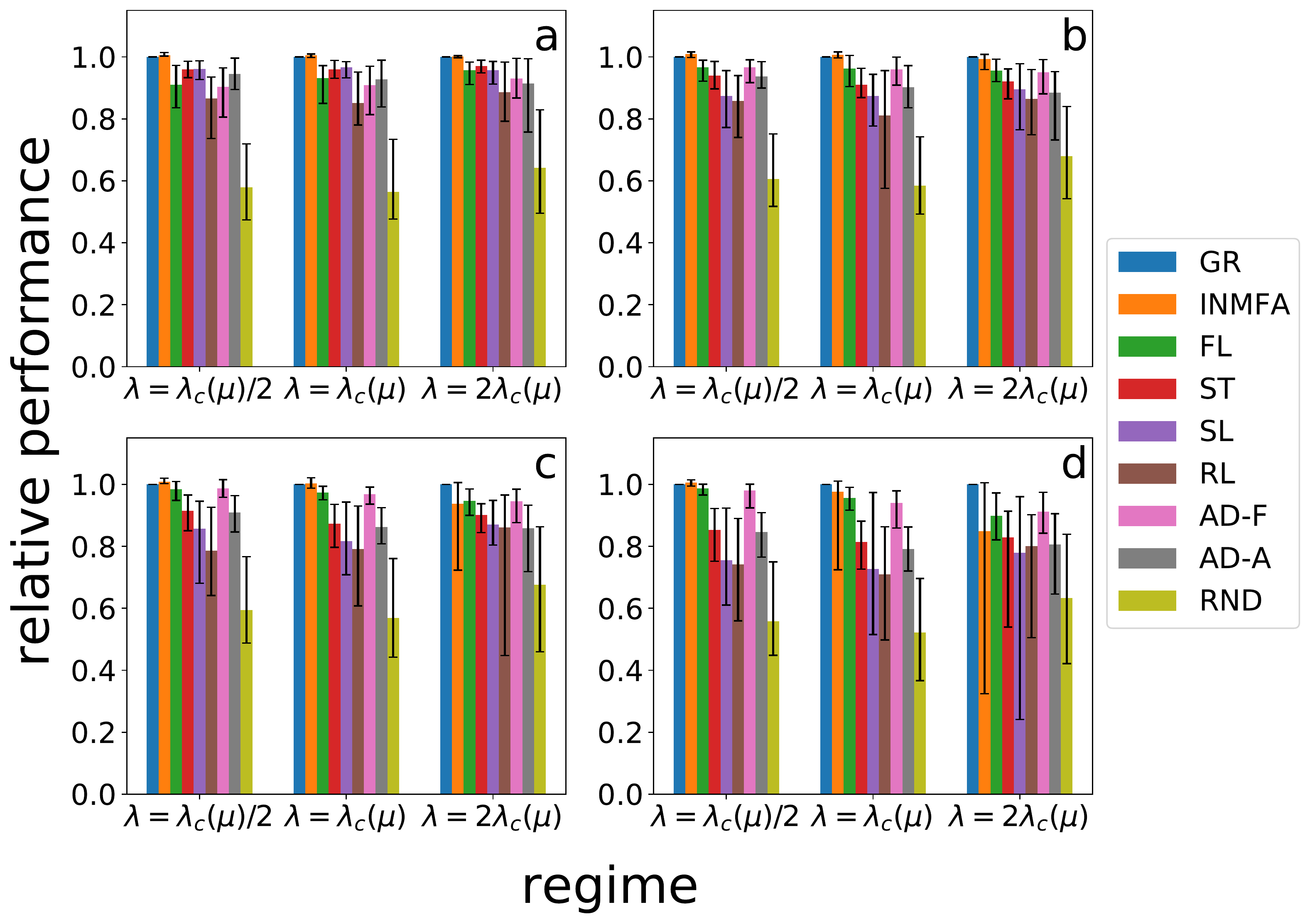}
\caption{{\bf Identification of influential spreaders in temporal networks.} (a) Performance, as defined in Eq.~(\ref{eq:performance}), of the various approximations listed in Table~\ref{table:methods}. Performance values are relative to those obtained for GR. The height of the colored bars indicate average values of the relative performance over the set of the twelve temporal networks studied in this paper, see Table~\ref{table:networks}. Error bars identify minimum and maximum values of the performance measured over the entire corpus of real networks. We study different dynamical regimes by selecting different spreading probability values while keeping the recovery probability fixed at $\mu = 0$. (b) Same as in (a), but for $\mu = 0.25$. (c) Same as in (a), but for $\mu = 0.5$. (d) Same as in (a), but for $\mu = 1$.}
\label{fig:5}
\end{figure}

In Fig.~\ref{fig:5}, we report summary results of our systematic analysis. Performance of the various approximations highly depend on both the parameters $\mu$ and $\lambda$. Many of the approximations reach  nearly optimal performances for small $\mu$ and $\lambda$ values. As $\mu$ grows, having perfect knowledge of the initial topology of the network, such as in the FL or AD-F approximations, becomes essential to reach good performance. Approximations that do not rely on such a knowledge lose $10-20\%$ in performance compared to the performance displayed by the same approximations at low $\mu$ values.

%%%%%%%%%%%%%%%%%%%%%%%%%%%%%%%%%%%%%%%%%%%%%%%%%%%%%%%%%%%%%%%%%%%%%%%%%%%%%%%%%%%%%%%%%%%%%%%%%%%%%%%%%%%%%%%%%%%%%%%%%%%%%%%%%%%%%%%%%%%%%%%%%%%%%%%%%%%%%%%%%%%%%%%%

\section{Conclusions}

Irreversible spreading models, such as the SIR model, display outcomes that strongly depend on the initial conditions. Such a sensitivity is apparent in static networks already, but gets amplified when the network exhibits temporal changes on a timescale comparable with the one of the spreading dynamics. While seeking solutions to the problem of influence maximization, sensitivity of the spreading outcome to initial conditions is further extremized. Indeed, our systematic analysis shows that good solutions to the influence maximization problem require accurate knowledge of the network dynamics, especially regarding the order in which network edges appear in the system. Also, one of our most important numerical findings is that having knowledge of only the first snapshot of a temporal network is still sufficient for identifying influential spreaders effectively. 
The topological characteristics of the nodes selected as spreaders depend on the dynamical regime. If the recovery probability $\mu$ is large, then nodes that are central in the first few layers are good spreaders. If $\mu$ is small instead, then nodes that are central on average over all layers are good spreaders.  For example, we see that AD-F outperforms AD-A in all settings except for the subcritical and critical regimes when $\mu=0$.

\begin{figure}[!htb]
\centering
\includegraphics[width=0.45\textwidth]{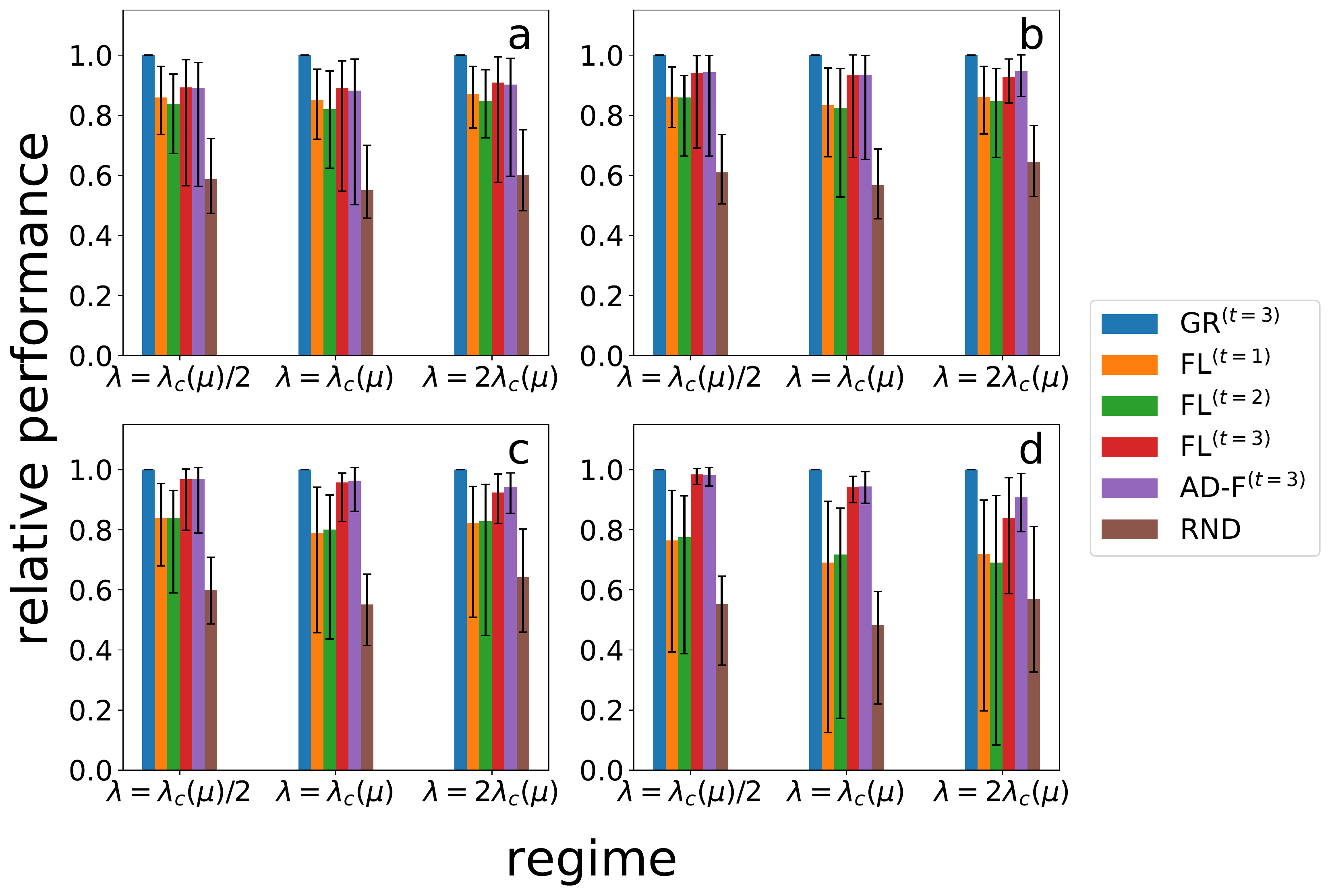}
\caption{{\bf Prediction of influential spreaders in temporal networks}. (a) Same as in Fig.~\ref{fig:5}(a) with the difference that the ground-truth dynamics is started at time $t=3$ instead of time $t=1$. Predictions using the $\textrm{GR}^{(t)}$, $\textrm{FL}^{(t)}$ and $\textrm{AD-F}^{(t)}$ approximations are based on perfect knowledge of the network topology/dynamics, but under the assumption that spreading starts at time $t$. (b) Same as in (a), but for $\mu = 0.25$. (c) Same as in (a), but for $\mu = 0.5$. (d) Same as in (a), but for $\mu = 1$.}
\label{fig:6}
\end{figure}

Our entire analysis is based on the evaluation of the performance of different ex-post approximations. In practical settings, however, it may be more realistic to expect the observer to be aware of past snapshots of a temporal network, and use this information to make predictions about top influencers for a future spreading process taking place on temporal network layers with unknown topology. As an illustrative example, in Fig.~\ref{fig:6}, we show performance results obtained by comparing ex ante predictions of the influential spreaders under approximations similar to those we considered earlier in the paper. Specifically, we name as $\textrm{FL}^{(t)}$ the approximation relying on layer $t$ as the only information available about the network, and with $\textrm{AD-F}^{(t)}$ the approximation relying on adaptive degree centrality computed on the $t$th layer of the network. We use layers $t=1, 2$ and $3$ to make predictions about the spreaders in the temporal network. The true dynamics starts from layer $t=3$ in the simulations.We measure the performance of our predictions relative to the best achievable one, here named as $\textrm{GR}^{(t=3)}$. The results of Fig.~\ref{fig:6} show that the lack of information about the initial layer of the dynamics leads to a significant drop in performance. Even a small delay between the last known layer and the start of the process may affect significantly the results. This fact indicates that further research is needed to design effective methods for the prediction of influential spreaders in temporal networks.

%%%%%%%%%%%%%%%%%%%%%%%%%%%%%%%%%%%%%%%%%%%%%%%%%%%%%%%%%%%%%%%%%%%%%%%%%%%%%%%%%%%%%%%%%%%%%%%%%%%%%%%%%%%%%%%%%%%%%%%%%%%%%%%%%%%%%%%%%%%%%%%%%%%%%%%%%%%%%%%%%%%%%%%%

\section*{Acknowledgement}

S. E. and F.R. acknowledge support from the National Science Foundation (CMMI-1552487). D.M. and F.R. acknowledge support from the US Army Research Office (W911NF- 16-1-0104).

%\bibliography{bibliography}

\begin{thebibliography}{10}

\bibitem{domingos2001mining}
P.~Domingos and M.~Richardson, ``Mining the network value of customers,'' in
  {\em Proceedings of the seventh ACM SIGKDD international conference on
  Knowledge discovery and data mining}, pp.~57--66, 2001.

\bibitem{kempe2003maximizing}
D.~Kempe, J.~Kleinberg, and {\'E}.~Tardos, ``Maximizing the spread of influence
  through a social network,'' in {\em Proceedings of the ninth ACM SIGKDD
  international conference on Knowledge discovery and data mining},
  pp.~137--146, 2003.

\bibitem{leskovec2007cost}
J.~Leskovec, A.~Krause, C.~Guestrin, C.~Faloutsos, J.~VanBriesen, and
  N.~Glance, ``Cost-effective outbreak detection in networks,'' in {\em
  Proceedings of the 13th ACM SIGKDD international conference on Knowledge
  discovery and data mining}, pp.~420--429, 2007.

\bibitem{chen2009efficient}
W.~Chen, Y.~Wang, and S.~Yang, ``Efficient influence maximization in social
  networks,'' in {\em Proceedings of the 15th ACM SIGKDD international
  conference on Knowledge discovery and data mining}, pp.~199--208, 2009.

\bibitem{goyal2011celf++}
A.~Goyal, W.~Lu, and L.~V. Lakshmanan, ``Celf++ optimizing the greedy algorithm
  for influence maximization in social networks,'' in {\em Proceedings of the
  20th international conference companion on World wide web}, pp.~47--48, 2011.

\bibitem{nguyen2016stop}
H.~T. Nguyen, M.~T. Thai, and T.~N. Dinh, ``Stop-and-stare: Optimal sampling
  algorithms for viral marketing in billion-scale networks,'' in {\em
  Proceedings of the 2016 International Conference on Management of Data},
  pp.~695--710, 2016.

\bibitem{hu2018local}
Y.~Hu, S.~Ji, Y.~Jin, L.~Feng, H.~E. Stanley, and S.~Havlin, ``Local structure
  can identify and quantify influential global spreaders in large scale social
  networks,'' {\em Proceedings of the National Academy of Sciences USA},
  vol.~115, no.~29, pp.~7468--7472, 2018.

\bibitem{estrada2005subgraph}
E.~Estrada and J.~A. Rodriguez-Velazquez, ``Subgraph centrality in complex
  networks,'' {\em Physical Review E}, vol.~71, no.~5, p.~056103, 2005.

\bibitem{kitsak2010identification}
M.~Kitsak, L.~K. Gallos, S.~Havlin, F.~Liljeros, L.~Muchnik, H.~E. Stanley, and
  H.~A. Makse, ``Identification of influential spreaders in complex networks,''
  {\em Nature physics}, vol.~6, no.~11, pp.~888--893, 2010.

\bibitem{de2014role}
G.~F. De~Arruda, A.~L. Barbieri, P.~M. Rodr{\'\i}guez, F.~A. Rodrigues,
  Y.~Moreno, and L.~da~Fontoura~Costa, ``Role of centrality for the
  identification of influential spreaders in complex networks,'' {\em Physical
  Review E}, vol.~90, no.~3, p.~032812, 2014.

\bibitem{morone2015influence}
F.~Morone and H.~A. Makse, ``Influence maximization in complex networks through
  optimal percolation,'' {\em Nature}, vol.~524, no.~7563, pp.~65--68, 2015.

\bibitem{lu2016vital}
L.~L{\"u}, D.~Chen, X.-L. Ren, Q.-M. Zhang, Y.-C. Zhang, and T.~Zhou, ``Vital
  nodes identification in complex networks,'' {\em Physics Reports}, vol.~650,
  pp.~1--63, 2016.

\bibitem{clusella2016immunization}
P.~Clusella, P.~Grassberger, F.~J. P{\'e}rez-Reche, and A.~Politi,
  ``Immunization and targeted destruction of networks using explosive
  percolation,'' {\em Physical review letters}, vol.~117, no.~20, p.~208301,
  2016.

\bibitem{zdeborova2016fast}
L.~Zdeborov{\'a}, P.~Zhang, and H.-J. Zhou, ``Fast and simple decycling and
  dismantling of networks,'' {\em Scientific reports}, vol.~6, p.~37954, 2016.

\bibitem{erkol2019systematic}
{\c{S}}.~Erkol, C.~Castellano, and F.~Radicchi, ``Systematic comparison between
  methods for the detection of influential spreaders in complex networks,''
  {\em Scientific reports}, vol.~9, no.~1, pp.~1--11, 2019.

\bibitem{holme2012temporal}
P.~Holme and J.~Saram{\"a}ki, ``Temporal networks,'' {\em Physics reports},
  vol.~519, no.~3, pp.~97--125, 2012.

\bibitem{prakash2010virus}
B.~A. Prakash, H.~Tong, N.~Valler, M.~Faloutsos, and C.~Faloutsos, ``Virus
  propagation on time-varying networks: Theory and immunization algorithms,''
  in {\em Joint European Conference on Machine Learning and Knowledge Discovery
  in Databases}, pp.~99--114, Springer, 2010.

\bibitem{karsai2011small}
M.~Karsai, M.~Kivel{\"a}, R.~K. Pan, K.~Kaski, J.~Kert{\'e}sz, A.-L.
  Barab{\'a}si, and J.~Saram{\"a}ki, ``Small but slow world: How network
  topology and burstiness slow down spreading,'' {\em Physical Review E},
  vol.~83, no.~2, p.~025102, 2011.

\bibitem{perra2012activity}
N.~Perra, B.~Gon{\c{c}}alves, R.~Pastor-Satorras, and A.~Vespignani, ``Activity
  driven modeling of time varying networks,'' {\em Scientific reports}, vol.~2,
  p.~469, 2012.

\bibitem{valdano2015analytical}
E.~Valdano, L.~Ferreri, C.~Poletto, and V.~Colizza, ``Analytical computation of
  the epidemic threshold on temporal networks,'' {\em Physical Review X},
  vol.~5, no.~2, p.~021005, 2015.

\bibitem{habiba2007maximizing}
T.~Habiba and T.~Berger-Wolf, ``Maximizing the extent of spread in a dynamic
  network,'' {\em Technical Report}, vol.~20, 2007.

\bibitem{osawa2015selecting}
S.~Osawa and T.~Murata, ``Selecting seed nodes for influence maximization in
  dynamic networks,'' in {\em Complex Networks VI}, pp.~91--98, Springer, 2015.

\bibitem{michalski2014seed}
R.~Michalski, T.~Kajdanowicz, P.~Br{\'o}dka, and P.~Kazienko, ``Seed selection
  for spread of influence in social networks: Temporal vs. static approach,''
  {\em New Generation Computing}, vol.~32, no.~3-4, pp.~213--235, 2014.

\bibitem{murata2018extended}
T.~Murata and H.~Koga, ``Extended methods for influence maximization in dynamic
  networks,'' {\em Computational social networks}, vol.~5, no.~1, p.~8, 2018.

\bibitem{han2017influence}
M.~Han, M.~Yan, Z.~Cai, Y.~Li, X.~Cai, and J.~Yu, ``Influence maximization by
  probing partial communities in dynamic online social networks,'' {\em
  Transactions on Emerging Telecommunications Technologies}, vol.~28, no.~4,
  p.~e3054, 2017.

\bibitem{zhuang2013influence}
H.~Zhuang, Y.~Sun, J.~Tang, J.~Zhang, and X.~Sun, ``Influence maximization in
  dynamic social networks,'' in {\em 2013 IEEE 13th International Conference on
  Data Mining}, pp.~1313--1318, IEEE, 2013.

\bibitem{gayraud2015diffusion}
N.~T. Gayraud, E.~Pitoura, and P.~Tsaparas, ``Diffusion maximization in
  evolving social networks,'' in {\em Proceedings of the 2015 acm on conference
  on online social networks}, pp.~125--135, 2015.

\bibitem{pastor2015epidemic}
R.~Pastor-Satorras, C.~Castellano, P.~Van~Mieghem, and A.~Vespignani,
  ``Epidemic processes in complex networks,'' {\em Reviews of modern physics},
  vol.~87, no.~3, p.~925, 2015.

\bibitem{erkol2018influence}
{\c{S}}.~Erkol, A.~Faqeeh, and F.~Radicchi, ``Influence maximization in noisy
  networks,'' {\em EPL (Europhysics Letters)}, vol.~123, no.~5, p.~58007, 2018.

\bibitem{gemmetto2014mitigation}
V.~Gemmetto, A.~Barrat, and C.~Cattuto, ``Mitigation of infectious disease at
  school: targeted class closure vs school closure,'' {\em BMC infectious
  diseases}, vol.~14, no.~1, p.~695, 2014.

\bibitem{stehle2011high}
J.~Stehl{\'e}, N.~Voirin, A.~Barrat, C.~Cattuto, L.~Isella, J.-F. Pinton,
  M.~Quaggiotto, W.~Van~den Broeck, C.~R{\'e}gis, B.~Lina, {\em et~al.},
  ``High-resolution measurements of face-to-face contact patterns in a primary
  school,'' {\em PloS one}, vol.~6, no.~8, 2011.

\bibitem{isella2011s}
L.~Isella, J.~Stehl{\'e}, A.~Barrat, C.~Cattuto, J.-F. Pinton, and W.~Van~den
  Broeck, ``What's in a crowd? analysis of face-to-face behavioral networks,''
  {\em Journal of theoretical biology}, vol.~271, no.~1, pp.~166--180, 2011.

\bibitem{vanhems2013estimating}
P.~Vanhems, A.~Barrat, C.~Cattuto, J.-F. Pinton, N.~Khanafer, C.~R{\'e}gis,
  B.-a. Kim, B.~Comte, and N.~Voirin, ``Estimating potential infection
  transmission routes in hospital wards using wearable proximity sensors,''
  {\em PloS one}, vol.~8, no.~9, 2013.

\bibitem{paranjape2017motifs}
A.~Paranjape, A.~R. Benson, and J.~Leskovec, ``Motifs in temporal networks,''
  in {\em Proceedings of the Tenth ACM International Conference on Web Search
  and Data Mining}, pp.~601--610, 2017.

\bibitem{mucha2010community}
P.~J. Mucha, T.~Richardson, K.~Macon, M.~A. Porter, and J.-P. Onnela,
  ``Community structure in time-dependent, multiscale, and multiplex
  networks,'' {\em Science}, vol.~328, no.~5980, pp.~876--878, 2010.

\bibitem{kivela2014multilayer}
M.~Kivel{\"a}, A.~Arenas, M.~Barthelemy, J.~P. Gleeson, Y.~Moreno, and M.~A.
  Porter, ``Multilayer networks,'' {\em Journal of complex networks}, vol.~2,
  no.~3, pp.~203--271, 2014.

\bibitem{SMat} See Supplemental Material at \url{http://homes.sice.indiana.edu/filiradi/Mypapers/Influence_maximization_on_temporal_networks_SM.pdf}
  
\bibitem{fournet2014contact}
J.~Fournet and A.~Barrat, ``Contact patterns among high school students,'' {\em
  PloS one}, vol.~9, no.~9, 2014.

\bibitem{mastrandrea2015contact}
R.~Mastrandrea, J.~Fournet, and A.~Barrat, ``Contact patterns in a high school:
  a comparison between data collected using wearable sensors, contact diaries
  and friendship surveys,'' {\em PloS one}, vol.~10, no.~9, 2015.

\bibitem{genois2015data}
M.~G{\'e}nois, C.~L. Vestergaard, J.~Fournet, A.~Panisson, I.~Bonmarin, and
  A.~Barrat, ``Data on face-to-face contacts in an office building suggest a
  low-cost vaccination strategy based on community linkers,'' {\em Network
  Science}, vol.~3, no.~3, pp.~326--347, 2015.

\bibitem{genois2018can}
M.~G{\'e}nois and A.~Barrat, ``Can co-location be used as a proxy for
  face-to-face contacts?,'' {\em EPJ Data Science}, vol.~7, no.~1, p.~11, 2018.

\bibitem{nemhauser1978analysis}
G.~L. Nemhauser, L.~A. Wolsey, and M.~L. Fisher, ``An analysis of
  approximations for maximizing submodular set functions—i,'' {\em
  Mathematical programming}, vol.~14, no.~1, pp.~265--294, 1978.

\end{thebibliography}
%\bibliographystyle{ieeetr}

\end{document}